\documentclass[letterpaper,english,reprint, aps,prb]{revtex4-2}
\usepackage[T1]{fontenc}
\usepackage[latin9]{inputenc}
\usepackage{amsmath}
\usepackage{amssymb}
\usepackage{xcolor}
\usepackage{dcolumn}
\usepackage{bm}
\usepackage{bbm}
\usepackage{braket}
\usepackage{mathdots}
\usepackage[caption=false]{subfig}
\usepackage{overpic}

\usepackage{multirow}
\usepackage{pifont}
\usepackage{bbding}
\usepackage{comment}
\usepackage{natbib}

\newcommand{\br}{\mathbf{r}}
\newcommand{\bq}{\mathbf{q}}

\makeatletter
\usepackage {youngtab} 
\usepackage{physics} 
\makeatother
\usepackage{babel}
\usepackage{hyperref}
\begin{document}
\preprint{APS/123-QED}
\title{Defect-Mediated Melting of Square-Lattice Solids
}
\author{William Grampel}
\affiliation{Department of Physics, Technion -- Israel Institute of Technology, Haifa 32000, Israel}

\author{Daniel Podolsky}
\email{podolsky@physics.technion.ac.il}

\affiliation{Department of Physics, Technion -- Israel Institute of Technology, Haifa 32000, Israel}
\date{\today}
\begin{abstract}
The Kosterlitz-Thouless-Halperin-Nelson-Young (KTHNY) theory successfully explains the melting mechanism of two-dimensional isotropic lattices as a two-step process
driven by the unbinding of topological defects. By considering the elastic theory of the square lattice, we extend the KTHNY theory to melting of square lattice solids. In addition to the familiar elastic constants that govern the theory -- the Young's modulus and the Poisson ratio -- a third constant controlling the anisotropy of the medium emerges.  This modifies both the logarithmic and angular interactions between the topological defects.  Despite this modification, the extended theory retains the qualitative features of the isotropic case, predicting a two-step melting with an intermediate tetratic phase. However, some subtle differences arise, including a modified bound on the translational correlation exponent and the absence of universal values for the Young's modulus at the solid-to-tetratic phase transition.
\end{abstract}
\maketitle
\section{Introduction}

In three dimensions, the melting of a solid to a liquid occurs through a direct first-order transition.  The solid breaks translational and orientational symmetries.  Both symmetries are restored simultaneously at the transition.

In two dimensions, an additional possibility exists: melting can proceed via a two-step process described by Kosterlitz, Thouless, Halperin, Nelson, and Young (KTHNY theory) \cite{Kosterlitz1973,PhysRevB.19.1855,PhysRevB.19.2457}. This mechanism involves the unbinding of topological defects: dislocations at a lower temperature, followed by disclinations at a higher one. The intermediate phase, in which translational order is lost but algebraic orientational order remains, is known as the hexatic or tetratic phase for triangular or square lattices, respectively.

KTHNY theory was originally formulated for systems with hexagonal symmetry, where only two elastic constants are needed. The theory has been thoroughly validated in such systems through both simulations and experiments, confirming the existence of an intermediate hexatic phase \cite{PhysRevLett.62.2401,PhysRevLett.58.1200,PhysRevLett.118.158001,PhysRevE.59.2594,PhysRevE.73.065104}.

In contrast, square lattices have three independent elastic constants, making their melting behavior more complex. The original works of KTHNY addressed some aspects of this case, but a complete theory remains undeveloped, partly due to the lack of concrete physical realizations.  However, recent numerical \cite{article,Abutbul_2022,PhysRevResearch.7.L012034} and experimental \cite{PhysRevE.76.040401,Walsh_2016,D4SM01377H} observations of tetratic phases in systems with 4-fold symmetry motivate a systematic study of square-lattice melting.

In this work, we generalize KTHNY theory to the square lattice by considering its full elastic description with three independent elastic constants: the Young's modulus, Poisson ratio, and an anisotropy parameter. Our main findings are:

(i) In the solid phase, the translational correlations decay algebraically, with a power $\eta_0$ whose upper bound is $\eta_0<\frac{1}{4}$.  This is a stricter bound than for the triangular lattice, where $\eta_0<\frac{1}{3}$.

(ii)  The interactions between dislocations in the solid phase depend on a specific combination of the Young's modulus and anisotropy parameter.  This combination flows to a universal value at the solid-to-tetratic transition; the remaining parameters remain non-universal.

(iii) In the tetratic phase, the orientational correlations decay algebraically with a power $\eta_4$.  This power is bounded by $\eta_4\le \frac{1}{4}$, which is the value obtained at the tetratic-to-liquid transition.  This matches the bound at the hexatic-to-liquid transition \cite{PhysRevB.19.2457}.

This article is organized as follows.  Section \ref{sec:elastic} reviews the elastic theory of the square lattice. Section \ref{sec:dislocham} derives the interaction between dislocations. Section \ref{sec:Renorm} develops the renormalization group (RG) equations, comparing them with the triangular case.  Section \ref{sec:crit} computes the decay exponents for the translational correlations in the solid phase, $\eta_0$, and for the orientational correlations in the tetratic phase, $\eta_4$.

\section{The elastic theory of the square lattice}
\label{sec:elastic}

In this section, we briefly review the linear elastic theory of a square-lattice solid.  The degrees of freedom are the two-component displacement field $u_{i}$ ($i\in\{x,y\}$).
The free energy is $F=\int d^2\br\, f(\br)$, with free energy density
\begin{align}
f(\br)=&\frac{1}{2}\big[\lambda(u_{xx}+u_{yy})^{2}+2\mu(u_{xx}^{2}+u_{yy}^{2}+2u_{xy}^{2})\nonumber\\ &\qquad+\gamma(u_{xx}^{2}+u_{yy}^{2})\big]\,.
\label{eq:f}
\end{align}
Here, $u_{ij}=\frac{\partial_{i}u_{j}+\partial_{j}u_{i}}{2}$ is the
symmetric strain, $\mu$ is the shear modulus, $\lambda$ is Lam\'e's first parameter, and $\gamma$ is an elastic constant which measures the axial anisotropy in the square lattice.  By comparison, for a triangular lattice, $\gamma=0$, corresponding to an isotropic medium.

The stress field is obtained from the variation of the free energy density,
\begin{align}
\sigma_{ij}&=\frac{\partial f}{\partial u_{ij}}
\end{align}
and its components are
\begin{align}
\sigma_{xx}&=(2\mu+\lambda+\gamma)u_{xx}+\lambda u_{yy}\nonumber\\
\sigma_{yy}&=(2\mu+\lambda+\gamma)u_{yy}+\lambda u_{xx}\\
\sigma_{xy}&=2\mu u_{xy}\nonumber
\end{align}
This is summarized by the elasticity tensor $M$, defined by
\begin{align}
\left(\begin{array}{c}
\sigma_{xx}\\
\sigma_{yy}\\
\sigma_{xy}
\end{array}\right)=M\left(\begin{array}{c}
u_{xx}\\
u_{yy}\\
2u_{xy}
\end{array}\right)
\end{align}
where 
\begin{align}
M=\left(\begin{array}{ccc}
2\mu+\lambda+\gamma & \lambda & 0\\
\lambda & 2\mu+\lambda+\gamma & 0\\
0 & 0 & \mu
\end{array}\right)\,.
\label{eq:K}
\end{align}
Thermodynamic stability requires $M$ to be positive definite.  This gives constraints on the elastic constants, 
\begin{align}
&\mu>0\,,\nonumber\\
&2\mu+\gamma>0\,,\label{eq:mu}\\
&2\mu+2\lambda+\gamma>0\,.\nonumber
\end{align}
Combining the last two equations, we also see that 
\begin{align}
2\mu+\lambda+\gamma>0\,.
\end{align}
We will find it convenient to re-express the elastic properties of a medium in terms of the Young's modulus, $Y$ and the Poisson ratio $\sigma$, and a dimensionless anisotropy parameter, $\zeta$, to be defined below. 
The Young's modulus and the Poisson ratio are defined through the compliance tensor
\begin{align}
\left(\begin{array}{c}
u_{xx}\\
u_{yy}\\
2u_{xy}
\end{array}\right)=\left(\begin{array}{ccc}
\frac{1}{Y} & -\frac{\sigma}{Y} & 0\\
-\frac{\sigma}{Y} & \frac{1}{Y} & 0\\
0 & 0 & \frac{1}{\mu}
\end{array}\right)\left(\begin{array}{c}
\sigma_{xx}\\
\sigma_{yy}\\
\sigma_{xy}
\end{array}\right)\,, \label{eq:comp}
\end{align}
which is the inverse of the elasticity tensor.  By inverting Eq.~\eqref{eq:K} we obtain
\begin{align}
Y&=\frac{(2\mu+\gamma)(2\mu+2\lambda+\gamma)}{2\mu+\lambda+\gamma} \label{eq:Y}\\
\sigma
&=\frac{\lambda}{2\mu+\lambda+\gamma}\label{eq:sigma}  
\end{align}
Note that the elasticity tensor, in terms of
Young's modulus and the Poisson ratio, is
\begin{align}
M=\left(\begin{array}{ccc}
\frac{Y}{1-\sigma^{2}} & \frac{\sigma Y}{1-\sigma^{2}} & 0\\
\frac{\sigma Y}{1-\sigma^{2}} & \frac{Y}{1-\sigma^{2}} & 0\\
0 & 0 & \mu
\end{array}\right)
\end{align}
Stability requires $-1<\sigma<1$ and $Y>0$.

In what follows, we will see that the parameter $\alpha$, defined by 
\begin{align}
\alpha\equiv\frac{1}{\mu}-\frac{2(1+\sigma)}{Y}=\frac{\gamma}{\mu(2\mu+\gamma)}\,,
\label{eq:alpha}
\end{align}
and its dimensionless counterpart 
\begin{align}
    \zeta=Y\alpha  \,\,,
\end{align}
are useful parameters to quantify the anisotropy of the system.
By using the thermodynamic stability conditions \eqref{eq:mu}, $\zeta$ satisfies
\begin{align}
\zeta>-4  \,\, .
\end{align}
Both parameters vanish in an isotropic medium.

\section{Hamiltonian for dislocations of a square lattice}\label{sec:dislocham}

In this section, we derive the interaction energy of dislocations on the square lattice.  For this, we first discuss the Airy stress function formalism and show how it is generalized to account for anisotropy. We then introduce dislocations as sources for the Airy stress function and use the formalism to derive the interaction between dislocations.

\subsection{Airy stress function}

In equilibrium, the stress fields satisfy 
\begin{align}
\partial_{j}\sigma_{ij}=0\,.    
\end{align}
Using the Vector Potential Theorem in two dimensions, there is a field $\chi_{i}$ that satifies
\begin{align}
\sigma_{ix}=\partial_{y}\chi_{i}\\
\sigma_{iy}=-\partial_{x}\chi_{i}\nonumber
\end{align}
Using the symmetry condition $\sigma_{yx}=\sigma_{xy}$ we also get
\begin{align}
\partial_{y}\chi_{y}=-\partial_x\chi_{x}
\end{align}
By using the Vector Potential Theorem again, 
the stress fields can be written as double derivatives of a scalar field, called the Airy stress function $\chi(\br)$:
\begin{align}
\sigma_{xy}=-\frac{\partial^{2}\chi}{\partial x\partial y},\,\sigma_{xx}=\frac{\partial^{2}\chi}{\partial^{2}y},\,\sigma_{yy}=\frac{\partial^{2}\chi}{\partial^{2}x}  \label{eq:stress-Airy} 
\end{align}
By the definition of the strain fields, they must satify the compatibility
condition,
\begin{align}
\frac{\partial^{2}u_{xx}}{\partial^{2}y}+\frac{\partial^{2}u_{yy}}{\partial^{2}x}-2\frac{\partial^{2}u_{xy}}{\partial x\partial y}=0\,,
\label{eq:compatibility}
\end{align}
where we have assumed that the displacement field is a smooth function of space.  Writing this condition in terms of $\chi$ gives
\begin{align}
\frac{1}{Y}\nabla^{4}\chi+\left(\frac{1}{\mu}-\frac{2+2\sigma}{Y}\right)\frac{\partial^{4}\chi}{\partial^{2}x\partial^{2}y}=0    
\end{align}
where $\nabla^4=(\partial_x^2+\partial_y^2)^2$ is the square of the Laplacian.
The term in parenthesis  is the parameter $\alpha$ defined in Eq.~\eqref{eq:alpha}. Hence, 
\begin{align}
\frac{1}{Y}\nabla^{4}\chi+\alpha\frac{\partial^{4}\chi}{\partial^{2}x\partial^{2}y}=0    
\end{align}
Note that for $\gamma=0$,
\begin{align}
\frac{1}{Y}\nabla^{4}\chi=0  \qquad (\mathrm{for}\,\,\,\gamma=0)\,,
\end{align}
{\em i.e.} for an isotropic material $\chi$ is a biharmonic function.

In the presence of dislocations, we allow the displacement field to become multivalued. Then, Eq.~\eqref{eq:compatibility} is replaced by
\begin{align}
\frac{\partial^{2}u_{xx}}{\partial^{2}y}+\frac{\partial^{2}u_{yy}}{\partial^{2}x}-2\frac{\partial^{2}u_{xy}}{\partial x\partial y}=\epsilon_{jk}\partial_{j}b_{k} \,,
\label{eq:compatibility_withDislocations}
\end{align}
where ${\bf b}(\br)$ is the Burgers vector field. This implies
\begin{align}
\frac{1}{Y}\left(\nabla^{4}\chi+\zeta\frac{\partial^{4}\chi}{\partial^{2}x\partial^{2}y}\right)=\epsilon_{jk}\partial_{j}b_{k}\,,  
\end{align}
which indicates that dislocations act as sources for the Airy stress function \cite{cha95},

\subsection{The dislocations Hamiltonian}

To derive the interaction between dislocations, it is useful to work in Fourier space,
\begin{align}
\frac{1}{Y}\left( q^{4}+\zeta q_{x}^{2}q_{y}^{2}\right)\widetilde{\chi}(\bq)=\widetilde{S}(\bq)\,,
\end{align}
where $q^4=({\bf q}^2)^2$ and $\widetilde{S}(\bq)=-i\epsilon_{jk}q_j\tilde{b}_k(\bq)$.  This yields
\begin{align}
\widetilde{\chi}(\bq)=Y\frac{\widetilde{S}(\bq)}{q^{4}+\zeta q_{x}^{2}q_{y}^{2}} \,.\label{eq:chi}
\end{align}
The free energy is $F=\int\frac{d^{2}\bq}{(2\pi)^{2}}\tilde{f}_\mathrm{disloc}(\bq)$, where
\begin{align}
\tilde{f}_\mathrm{disloc}(\bq)=\frac{1}{2}\sum_{jk}\widetilde{u}_{jk}(\bq)\widetilde{\sigma}_{jk}(-\bq)
\end{align}
is the free energy density. By using Eq.~\eqref{eq:stress-Airy}, \eqref{eq:chi}, and \eqref{eq:comp} we arrive at 
\begin{align}
\tilde{f}_\mathrm{disloc}(\bq)=\frac{Y}{2}\frac{\widetilde{S}(\bq)\widetilde{S}(-\bq)}{q^{4}+\zeta q_{x}^{2}q_{y}^{2}}
\end{align}
with 
\begin{align}
\widetilde{S}(\bq)\widetilde{S}(-\bq)=(q^{2}\delta_{ij}-q_{i}q_{j})\widetilde{b}_{i}(\bq)\widetilde{b}_{j}(-\bq)    
\end{align}
For a collection of dislocations, the Burgers vector field can be written as
\begin{align}
b_{i}({\bf r})&=\sum_{\ell=1} b_{i}^\ell\delta({\bf r}-{\bf R}_\ell)
\end{align}
where ${\bf b}^\ell$ is the Burgers vector of the dislocation located at ${\bf R}_\ell$.  The Fourier transform is:
\begin{align}
\widetilde{b}_{i}(\bq)&=\sum_\ell b_{i}^\ell e^{i{\bf R}_\ell\cdot \bq}\nonumber  
\end{align}
The energy of the dislocation field is then:
\begin{align}
E=\sum_{\ell_1,\ell_2}\frac{Y}{2}\int\frac{d^{2}{\bf q}}{(2\pi)^{2}}\frac{b_{i}^{\ell_1}b_{j}^{\ell_2}(q^{2}\delta_{ij}-q_{i}q_{j}) e^{i({\bf R}_{\ell_1}-{\bf R}_{\ell_2}) \cdot \bq}}{q^{4}+\zeta q_{x}^{2}q_{y}^{2}}    \label{eq:dislocE}
\end{align}


We'll consider the energy of two dislocations separated by a distance ${\bf R}={\bf R}_2-{\bf R}_1$ with opposite Burgers vectors in the $x$-direction, ${\bf b}^1=a_0 \hat{x}=-{\bf b}^2$, where $a_0$ is the lattice spacing. We consider a general separation between dislocations ${\bf R}$ 
\begin{align}
    {\bf R} =R(\cos\phi,\sin\phi)
\end{align}
where $\phi$ is the angle of ${\bf R}$ relative to the $x$ axis. The integral in Eq.~\eqref{eq:dislocE} is then
\begin{align}
E&=\frac{Y a_0^2}{2}\int\frac{d^{2}{\bf q}}{(2\pi)^{2}}\frac{q^{2}_y\,\left(2-2e^{i{\bf R}\cdot \bq}\right)}{q^{4}+\zeta q_{x}^{2}q_{y}^{2}}  \label{eq:twodisloc}\\&=\frac{Y a_0^2}{2}\int\frac{d^{2}{\bf q}}{(2\pi)^{2}}\frac{g(\theta)}{q^2}\left(2-2e^{i\cos(\theta-\phi) Rq }\right)\nonumber
\end{align}
where we wrote $q_x=q\cos\theta,q_y=q\sin\theta$, and introduced
\begin{align}
g(\theta)=\frac{q^{2}_y}{q^2+\zeta q_{x}^{2}q_{y}^{2}/q^2}=\frac{\sin^2\theta}{1+\zeta \cos^2\theta\sin^2\theta}\,.
\end{align}
It is difficult to evaluate Eq.~(\ref{eq:twodisloc}) in closed form.  

Progress can be made by decomposing $g(\theta)$, which is an even and periodic function of $\theta$ with period $\pi$, into a sum of  harmonics of $2\theta$ 
\begin{align} 
g(\theta)=\sum_{n=0}{h_{2n} \cos(2n\theta)}\,. \label{eq:harmonic}
\end{align}
For example, the first three harmonics are
\begin{align} 
h_{0}&=\int_0^{2\pi}\frac{d\theta}{2\pi}g(\theta)=\frac{1}{2\sqrt{1+\frac{\zeta}{4}}}\, , \label{eq:harmonic2}\\
h_{2}&=\int_0^{2\pi}\frac{d\theta}{2\pi}2\cos(2\theta) g(\theta)=-\frac{4\sqrt{1+\frac{\zeta}{4}}-4}{\zeta}\, ,\nonumber
\\h_4&=\int_0^{2\pi}\frac{d\theta}{2\pi}2\cos(4\theta) g(\theta)=\frac{8+\zeta-8\sqrt{1+\frac{\zeta}{4}}}{\zeta\sqrt{1+\frac{\zeta}{4}}}\nonumber
\end{align}
In particular, for an isotropic medium $(\zeta=0)$ $h_0=-h_2=1/2$ and all the higher harmonics vanish. 

The derivation of the interaction appears in Appendix \ref{sec:AppDerivationOfInteraction}.
We obtain 
\begin{align}
\label{eq:disloc_energy}
E= \frac{K a_0^2}{4\pi}\ln \frac{R}{a}+W(\phi)
\end{align}
where
\begin{align}
    K=\frac{Y}{\sqrt{1+\frac{\zeta}{4}}}
    \label{eq:KY}
\end{align}
controls the logarithmic interaction between dislocations
and
\begin{align}
W(\phi)=-Y a_0^2\sum_{n=1}(-1)^n\frac{h_{2n}(\zeta)}{4\pi n}\cos(2n\phi)
\end{align}
is the anisotropic part of the interaction, which depends only on the angle of the separation vector, $\phi$, but not on its magnitude, $R$.

\begin{figure}
\begin{overpic}[width=0.4\textwidth]{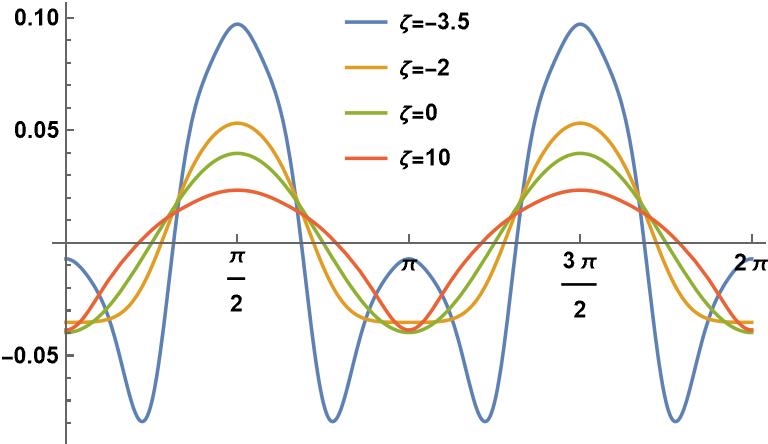}
  \put(3, 62){\Large{$\frac{W}{Ya_0^2}$}}
  \put(100, 30){\Large{$\phi$}}
 \end{overpic}
   \caption{Angular dependence of the interaction energy between dislocations of opposite Burgers vectors, $W(\phi)$, for different values of the anisotropy $\zeta$.}
  \label{angular}
\end{figure}

Figure \ref{angular} displays $W(\phi)$ for different values of $\zeta$.
For the isotropic case ($\zeta=0$), the minimal energy configuration is obtained for $\phi=0$ (mod $\pi$), i.e. when the separation between the dislocations is aligned with their Burgers vectors. 
It is interesting to note that if $\zeta$ is negative enough, the $n=2$ term becomes significant. Then, the minimal energy and most probable angle between the two dislocation is shifted away from $\phi=0$.  

The above results apply to two dislocations with Burgers vectors in the $\pm \hat{x}$ direction. For two dislocations in the $\pm\hat{y}$ direction, $\phi$ is shifted by $\frac{\pi}{2}$:
\begin{align}
\label{eq:disloc_energy2}
 E=\frac{K a_0^2}{4\pi}\ln \frac{R}{a}+W\left(\phi+\frac{\pi}{2}\right)\,,
\end{align}

By comparison, for $\zeta=0$, only the logarithmic dependence and the second harmonic survive, leading to the familiar isotropic interaction:
\begin{align}
E=\frac{Y}{4\pi }\left[\delta_{ij}\ln \frac{R}{a}-\frac{R_{i}R_{j}}{R^{2}}\right]b_{i}b_{j} \quad \mathrm{(isotropic)}
\label{eq:disloc_interaction}
\end{align}
where the logarithmic interaction between dislocations is controlled by the Young's modulus $Y$. 

\begin{figure}
\begin{overpic}[width=0.4\textwidth]{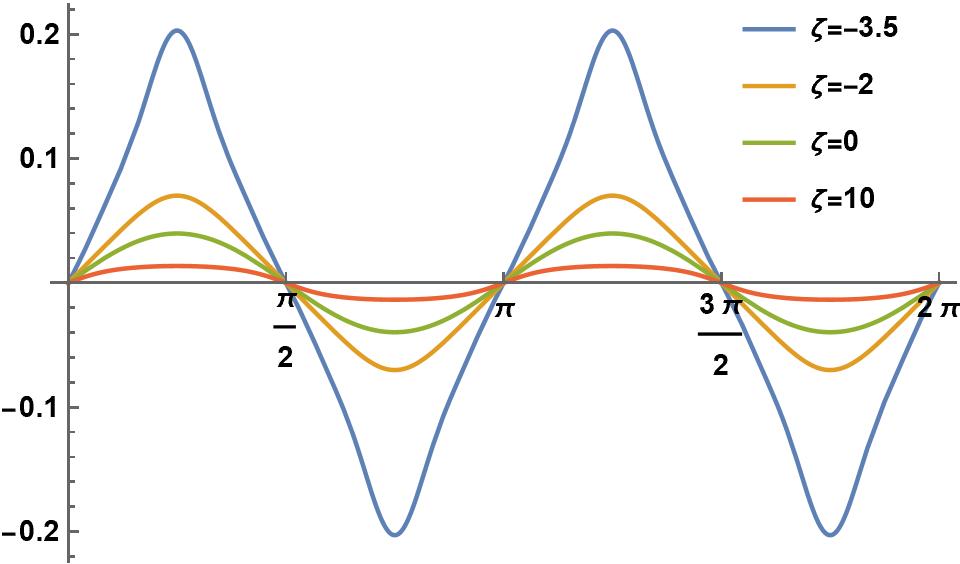}
  \put(3.1, 62){\Large{$\frac{\widetilde{W}}{Ya_0^2}$}}
  \put(100, 33){\Large{$\phi$}}
 \end{overpic}
   \caption{Angular dependence of the interaction energy of perpendicular dislocations, $\widetilde{W}(\phi)$, for different values of $\zeta$.}
  \label{angular perp}
\end{figure}

Finally, it is interesting to compute the interaction energy between dislocations with perpendicular Burgers vectors, ${\bf b}_1= a_0\hat{x}$ and ${\bf b}_1= a_0\hat{y}$.  In this case, the energy of the system diverges in the thermodynamic limit since the total Burgers vector of the system is non-zero.  However, if these dislocations are part of a neutral ensemble, then the energy remains finite, and we can compute the contribution of the interaction between this dislocation pair to the total energy.  This is derived in Appendix \ref{sec:AppDerivationOfInteraction}. The interaction energy between perpendicular dislocations, $\widetilde{W}(\phi)$, is found to be independent of their distance, and to depend only on the angle $\phi$.  The interaction energy is plotted in Figure \ref{angular perp}. The most probable angles are $\phi=\frac{3\pi}{4}$ and $\phi=\frac{7\pi}{4}$ regardless of $\zeta$.
\section{Renormalization of the square lattice}\label{sec:Renorm}

In this section, we develop the RG for the square lattice parameters.
To get the flow equations for the renormalized constants we will look
at the response function of the elastic tensor.

\subsection{The inverse elasticity tensor of the square lattice}

The renormalization of the elastic constants can be extracted from the renormalization of the inverse elastic tensor, as will be discussed in Section \ref{sec:RenormInverseElasticity}.  In this section, we will derive this tensor. 

The reduced free energy, written in tensor form, is
\begin{align}
\beta F=\frac{1}{2}\int\frac{d^{2}{\bf r}}{a_{0}^{2}}C_{ij,kl}u_{ij}u_{kl}    
\end{align}
where $u_{ij}$ is the symmetric strain, $a_{0}$ is the lattice spacing
and $C_{ijkl}$ is the tensor of reduced elastic constants which are related to the bare constants by:
\begin{align}
\mu\equiv\frac{\mu_{0}a_{0}^{2}}{T}, \lambda\equiv\frac{\lambda_{0}a_{0}^{2}}{T}, \gamma\equiv\frac{\gamma_{0}a_{0}^{2}}{T}    
\end{align} 
We also introduce the reduced interaction strength
\begin{align}
K\equiv\frac{K_{0}a_{0}^{2}}{T},    
\end{align}
For the square lattice, $C_{ij,kl}$ takes the form
\begin{align}
C_{ij,kl} & =\mu(\delta_{ik}\delta_{jl}+\delta_{il}\delta_{jk})+\lambda\delta_{ij}\delta_{kl}\\
 & +\gamma(\delta_{i1}\delta_{j1}\delta_{k1}\delta_{l1}+\delta_{i2}\delta_{j2}\delta_{k2}\delta_{l2})\nonumber
\end{align}
The inverse of this tensor is computed in App.~\ref{app:DerivationInverseTensor},
\begin{align}
C_{ij,kl}^{-1} & =\frac{1}{4\mu}(\delta_{ik}\delta_{jl}+\delta_{il}\delta_{jk})-\frac{\lambda}{(2\mu+\gamma)(2\mu+2\lambda+\gamma)}\delta_{ij}\delta_{kl}\nonumber\\
 & -\frac{\gamma}{2\mu(2\mu+\gamma)}(\delta_{i1}\delta_{j1}\delta_{k1}\delta_{l1}+\delta_{i2}\delta_{j2}\delta_{k2}\delta_{l2})
\end{align}
whose non-zero entries are:
\begin{align}
C_{yx,yx}^{-1}&=C_{xy,xy}^{-1}=\frac{1}{4\mu}\\
C_{xx,xx}^{-1}&=C_{yy,yy}^{-1}\nonumber\\
&=\frac{1}{2\mu}-\frac{\lambda}{(2\mu+\gamma)(2\mu+2\lambda+\gamma)}-\frac{\gamma}{2\mu(2\mu+\gamma)}\nonumber\\
C_{xx,yy}^{-1}&=C_{yy,xx}^{-1}=-\frac{\lambda}{(2\mu+\gamma)(2\mu+2\lambda+\gamma)}\nonumber    
\end{align}
\noindent From these we also get the relation
\begin{align}
4C_{yx,yx}^{-1}-2C_{xx,xx}^{-1}+2C_{xx,yy}^{-1}=\frac{\gamma}{\mu(2\mu+\gamma)}\equiv \alpha    
\label{eq:CtoAlpha}
\end{align}
which is the $\alpha$ parameter introduced in Eq.~(\ref{eq:alpha}).

\subsection{Renormalization of the inverse elasticity tensor }
\label{sec:RenormInverseElasticity}

The renormalization of the inverse elasticity tensor can be computed by expressing it as a response function  \cite{PhysRevB.19.1855}:
\begin{align}
C_{R,ijkl}^{-1}=\frac{1}{\Omega a_{0}^{2}}\left\langle U_{ij}U_{kl}\right\rangle     
\end{align}
where $\Omega$ is the area and $U_{ij}$ is the area-integrated strain,
\begin{align}
U_{ij}=\int d^2r\, u_{ij}\,.  
\end{align}
For a generic strain field $u_{ij}$, $U_{ij}$ decomposes as
\begin{align}
U_{ij}=\int d^{2}r \phi_{ij}+U_{ij}^\mathrm{sing}    
\end{align}
where $\phi_{ij}$ is the smooth part of $u_{ij}$ and where the singular part can be written in as 
\begin{align}
U_{ij}^{\mathrm{sing}}=\frac{1}{2}a_{0}\sum_{\bf R}\left[b_{i}({\bf R})\epsilon_{jk}R_{k}+b_{j}({\bf R})\epsilon_{ik}R_{k}\right] \,.
\end{align}
where the sum is over the locations ${\bf R}$ of the dislocations, and $b_i({\bf R})$ is the $i$th component of the Burgers vector at ${\bf R}$.

The inverse of the bare elasticity tensor can be written in terms of the smooth part as,
\begin{align}
C_{ijkl}^{-1}=\frac{1}{\Omega a_{0}^{2}}\left\langle \int d^{2}{\bf r}_{1}\phi_{ij}(r_{1})\int d^{2}{\bf r}_{2}\phi_{ij}(r_{2})\right\rangle     
\end{align}
This gets renormalized by the contribution of singular fields
\begin{align}
C_{R,ijkl}^{-1}=C_{ijkl}^{-1}+\frac{1}{\Omega a_{0}^{2}}\left\langle U_{ij}^\mathrm{sing}U_{kl}^\mathrm{sing}\right\rangle \end{align}
The cross terms of smooth and singular parts vanish upon spatial integration because they are not correlated in the linear elastic theory.

To compute the renormalization to second order in fugacity $y=e^{-\frac{E_{c}}{T}}$ with dislocation core energy $E_c$,
we consider the singular field for a neutral pair of fundamental dislocations $\pm{\bf b}$:
\begin{align}
U_{ij}^\mathrm{sing}=\frac{1}{2}a_{0}(b_{i}\epsilon_{jk}R_{k}+b_{j}\epsilon_{ik}R_{k})    
\end{align}
Here ${\bf R}$ is the separation between the pair. In a square lattice, $\bf{b}$ can be either $a_0 {\bf e}_{x}$ or $a_0 {\bf e}_{y}$. The renormalized values of the inverse tensor are then:
\begin{widetext}
\begin{align}
\label{eq:Crenorm}
C_{R,ijkl}^{-1} & =C_{ijkl}^{-1}+\frac{1}{4}y^{2}(e_{x,i}\epsilon_{js}+e_{x,j}\epsilon_{is})(e_{x,k}\epsilon_{lt}+e_{x,l}\epsilon_{kt})\int\frac{d^{2}{\bf R}}{a^{2}}\frac{R_{s}R_{t}}{a^{2}}e^{-W(\theta)}\left(\frac{R}{a}\right)^{-\frac{K}{4\pi}}\\
& +\frac{1}{4}y^{2}(e_{y,i}\epsilon_{js}+e_{y,j}\epsilon_{is})(e_{y,k}\epsilon_{lt}+e_{y,l}\epsilon_{kt})\int\frac{d^{2}{\bf R}}{a^{2}}\frac{R_{s}R_{t}}{a^{2}}e^{-W(\theta+\frac{\pi}{2})}\left(\frac{R}{a}\right)^{-\frac{K}{4\pi}}\nonumber
\end{align}
\end{widetext}
Here, $\theta$ is the angle of ${\bf R}$ relative to the $x$-axis and $a$ is the dislocation core radius, which also acts as the minimal separation between dislocations in the integration. Following Eqs. (\ref{eq:disloc_energy}) and (\ref{eq:disloc_energy2}), the factors  $e^{H(\theta)}\,(\frac{R}{a})^{-\frac{K}{4\pi}}$
and $e^{H(\theta+\pi/2)}\,(\frac{R}{a})^{-\frac{K}{4\pi}}$
are the statistical weights $e^{-\beta E}$ for the occurrence of pairs of dislocations.  The renormalization only includes integration over pairs of dislocations and not triplets of dislocations, as required in the treatment of the triangular lattice where three dislocations can form a neutral configuration. 

Equation ~\eqref{eq:Crenorm} gives the renormalization of the elastic constants to order $y^2$.  We obtain a flow equation for each of the three independent components, $C_{yx,yx}^{-1}$, $C_{xx,xx}^{-1}$, and $C_{xx,yy}^{-1}$.
Inspection of Eq.~\eqref{eq:Crenorm} shows that $C_{xx,yy}^{-1}$ is not renormalized by the dislocation pairs. Therefore,
\begin{align}
\frac{\lambda_{R}}{(2\mu_{R}+\gamma_{R})(2\mu_{R}+2\lambda_{R}+\gamma_{R})}&=-C_{R,xxyy}^{-1} \label{eq:constofflow}
\end{align}
equals its bare value $\frac{\lambda}{(2\mu+\gamma)(2\mu+2\lambda+\gamma)}$. 
From the other components we obtain the renormalization of $1/\mu$ and $\alpha$:
\begin{align}
\frac{1}{\mu_{R}}&=4C_{R,yxyx}^{-1}\\&=\frac{1}{\mu}+2y^{2}\int\frac{d^{2}{\bf R}}{a^{2}}\frac{R^{2}}{a^{2}}\text{cos}^{2}\theta e^{-W(\theta)}\left(\frac{R}{a}\right)^{-\frac{K}{4\pi}}\nonumber\\&=\frac{1}{\mu}+y^{2}\int\frac{d^{2}{\bf R}}{a^{2}}\frac{R^{2}}{a^{2}}\left(1+\cos2\theta\right) e^{-W(\theta)}\left(\frac{R}{a}\right)^{-\frac{K}{4\pi}}\nonumber
\end{align}
and, using Eq.~(\ref{eq:CtoAlpha}),
\begin{align}
\alpha_{R}&=4C_{R,yxyx}^{-1}-2C_{R,xxxx}^{-1}+2C_{R,xxyy}^{-1}\\&=\alpha+2y^{2}\int\frac{d^{2}{\bf R}}{a^{2}}\frac{R^{2}}{a^{2}}\left(\cos^{2}\theta -\sin^2\theta \right)e^{-W(\theta)}\left(\frac{R}{a}\right)^{-\frac{K}{4\pi}}\nonumber\\&=\alpha+2y^{2}\int\frac{d^{2}{\bf R}}{a^{2}}\frac{R^{2}}{a^{2}}\cos2\theta  e^{-W(\theta)}\left(\frac{R}{a}\right)^{-\frac{K}{4\pi}}\nonumber\label{eq:renorm}
\end{align}
We now define the angular intergrals:
\begin{align}
    Q_0(Y,\zeta)&=\int_0^{2\pi}d\theta e^{-W(\theta)}\\
    Q_1(Y,\zeta)&=\int_0^{2\pi}d\theta\cos2\theta e^{-W(\theta)}
\end{align}
where the dependence on $Y$ and $\zeta$ enters through $W(\theta)$.
Since $e^{-W(\theta)}> 0$ and $1\pm \cos 2\theta\ge0$, they satisfy
\begin{align}
    &Q_0(Y,\zeta)>0,\label{eq:angcondition1}\\  &Q_0(Y,\zeta)\pm Q_1(Y,\zeta)>0 \label{eq:angcondition2}
\end{align}
for all values of $Y$ and $\zeta$.
The renormalization equations then become
\begin{align}
\frac{1}{\mu_{R}}-\frac{1}{\mu}&= \left(Q_0+Q_1\right)y^{2} \int_a^{\infty}\frac{dR}{a}\left(\frac{R}{a}\right)^{3-\frac{K}{4\pi}}\label{eq:RenormMu}\\
\alpha_R-\alpha&=2Q_1y^{2}\int_a^{\infty}\frac{dR}{a}\left(\frac{R}{a}\right)^{3-\frac{K}{4\pi}}\label{eq:RenormAlpha} 
\end{align}
In what follows, we find that the RG flow is not sensitive to the precise form of the functions $Q_0(Y,\zeta),Q_1(Y,\zeta)$, provided they satisfy \eqref{eq:angcondition1} and \eqref{eq:angcondition2}.

\subsection{RG flow equations}
In order to obtain the renormalization group equations, we consider the renormalization due to dislocation pairs as we continuously integrate out short distances. We start by separating the integrals in Eqs.~\eqref{eq:RenormMu} and \eqref{eq:RenormAlpha} into two parts:
\begin{equation}
    \int_a^\infty=\int_a^{ae^{dl} }+\int_{ae^{dl} }^\infty
\end{equation}
By absorbing the first part into the bare values, followed by rescaling of the cuttoff
\begin{align}
ae^{dl}\rightarrow a
\end{align}
we get the same equations as in Eqs.~\eqref{eq:RenormMu} and \eqref{eq:RenormAlpha} but with shifted and rescaled bare parameters and fugacity. The fugacity scales as
\begin{align}
    y\rightarrow e^{dl(2-\frac{K}{8\pi})}y
\end{align}
We emphasize that the renormalization depends on the coupling $K$ and not the Young's modulus, as in the isotropic case. A continuous application of this procedure gives the flow equations. The flow for the fugacity explicitly depends only
on the coupling constant:
\begin{align}
\frac{d}{dl}y(l)=\left(2-\frac{K(l)}{8\pi}\right)y(l) 
\label{eq:yFlow}
\end{align}
It does not have a $O(y^{2})$ term that is present in the triangular case because only pairs of dislocations are considered. As expected, there is a phase transition at $K=16\pi$, which is reminiscent of the transition in the triangular lattice. However, unlike the triangular lattice, for the square lattice $K$ does not uniquely determine the Young's modulus at the transition. Instead, using Eq.~\eqref{eq:KY} we obtain that, at the transition, 
\begin{align}
\frac{Y}{\sqrt{1+\frac{\zeta}{4}}}=16\pi\,.    
\end{align}
Hence, only this specific combination of the Young's modulus $Y$ and the anisotropic parameter $\zeta$ is universal.  In addition, the Poisson ratio can take a range of values at the transition.

The rescaling of $\frac{1}{\mu}$ and $\alpha$ from Eqs.~\eqref{eq:RenormMu} and \eqref{eq:RenormAlpha} gives us the flow equations:
\begin{align}
&\frac{d}{dl}\frac{1}{\mu}=(Q_0+Q_1)y^{2}(l)\label{eq:muFlow}\\&\frac{d}{dl}\alpha=2Q_1y^{2}(l)\label{eq:alphaFlow}
\end{align}
In addition, from Eq.~\eqref{eq:constofflow} we obtain
\begin{align}
   \frac{d}{dl} \frac{\lambda}{(2\mu+\gamma)(2\mu+2\lambda+\gamma)}=0
   \label{eq:constofflow2}
\end{align}
is a constant of flow.  Equations  \eqref{eq:yFlow},  \eqref{eq:muFlow}, \eqref{eq:alphaFlow}, and \eqref{eq:constofflow2} fully determine the RG flow of the system.

From these equations, we also get equations of flow for other useful combinations of the elastic constants,
\begin{align}
    \frac{d}{dl}\frac{2}{2\mu+\gamma}=\frac{d}{dl}\left(\frac{1}{\mu}-\alpha\right)=(Q_0-Q_1)y^{2}\label{eq:anotherflow}(l)
\end{align}
From \eqref{eq:constofflow2} and \eqref{eq:anotherflow} we get the flow for $\lambda$ and $\frac{1}{Y}$:
\begin{align}
   \frac{d}{dl} \frac{1}{\lambda}&=y^{2}\frac{2\mu+\lambda+\gamma}{\lambda} (Q_0-Q_1)\\
    \frac{d}{dl}\frac{1}{Y}&=y^{2}\frac{Q_0-Q_1}{2} 
\end{align}
The equation for $\zeta=Y\alpha$ is then
\begin{align}
   \frac{d}{dl}\zeta
   =y^{2}Y\left(2Q_1+\zeta\frac{Q_1-Q_0}{2}\right)
\end{align}
Finally, from $K^{-1}=\frac{1}{Y}\sqrt{1+\frac{\zeta}{4}}$
we obtain 
\begin{align}
    \frac{d({K}^{-1})}{dl}
    =\frac{y^2}{4}\frac{(Q_0-Q_1)(2+\frac{\zeta}{4})+{Q_1}}{\sqrt{1+\frac{\zeta}{4}}}\,.
\end{align}

From the conditions \eqref{eq:angcondition1} and \eqref{eq:angcondition2}, the flow of $K^{-1}$ is in the positive direction and it only stops if the fugacity vanishes. The flows of the anisotropy $\zeta$ can change sign: if $\zeta$ starts small, or negative, it grows; if it starts sufficiently large, it decreases. As pointed out in \cite{PhysRevB.19.2457},  $\zeta$ will flow away from $0$ even if its bare value vanishes. This is because the Burgers vectors are constrained to be $\pm a_0{\bf e}_{x}$ and $\pm a_0{\bf e}_{y}$, which in itself introduces anisotropy to the medium.

\section{Critical exponents}\label{sec:crit}

 In this section, we compute the power-law exponents of the translational and orientational correlation functions in the solid and tetratic phases, respectively. 

\subsection{Translational critical exponent for the square lattice solid}

A two-dimensional solid has long-range orientational order that is accompanied by algebraic translational order.  
The order parameter of a broken translational symmetry is 
\begin{align}
\rho_G({\bf R})=e^{i\bf{G}\cdot\left({\bf R}+{\bf u}({\bf R})\right)} 
\end{align}
where  ${\bf G}$ is a reciprocal lattice vector. The translational correlation function is then
\begin{align}
    C_G({\bf R})=\langle\rho_G\left({\bf R} \right)\rho^*_G(0)\rangle
\end{align}
The decay of the correlation at large distances, which is characterized by an exponent $\eta_G$, 
\begin{align}
C_{G}({\bf R})\sim R^{-\eta_{G}}  
\end{align}
will be computed in this section.

\subsubsection{Calculation of the critical exponent}
The correlation function for $Q$ near a reciprocal lattice vector \cite{PhysRevB.19.2457}
\begin{align}
  C_{Q}(R)=\langle e^{i\bf{Q}\cdot[ \textbf{u}\left(\textbf{R} \right)-\textbf{u}\left(\textbf{0} \right)]}\rangle 
\end{align}
can be written by using the cummulant expansion as
\begin{align}
  C_{Q}(R)\approx e^{-\frac{1}{2}Q_iQ_j\langle[\textbf{u}_i\left(\textbf{R}\right)-\textbf{u}_i\left(\textbf{0}\right)][\textbf{u}_j\left(\textbf{R}\right)-\textbf{u}_j\left(\textbf{0} \right)]\rangle} 
\end{align}
which is correct to $O(y^2)$ for the dislocation pair part of the displacement field \cite{PhysRevB.19.2457}. By inserting the Fourier transform of $\textbf{u}(\textbf{R})$,
\begin{align}
    \textbf{u}(\textbf{q})=\sum_{\textbf{R}}e^{i\textbf{q}\cdot\textbf{R}}\textbf{u}(\textbf{R})
\end{align}
The translational correlation $C_{Q}(R)$ for $Q$ near a reciprocal lattice vector  can we written as  
\begin{align}
C_{Q}(R)=e^{Q_{i}Q_{j}G_{ij}(R)}    
\end{align}
with 
\begin{align}
G_{ij}(R)=\int_\bq\langle u_{i}(\textbf{q})u_{j}(-\textbf{q})\rangle(e^{{i\textbf{q}}\cdot\textbf{R}}-1)
\end{align}
where we introduced the short-hand notation $\int_\bq\equiv \int\frac{d^{2}\bq}{(2\pi)^2}$. 
We are interested in the behavior of $C_{Q}(R)$ in the large $R$ limit.
$\langle u_{i}(\textbf{q})u_{j}(-\textbf{q})\rangle$ is evaluated in the limit of small $q$ with the renormalized elastic constants
\begin{align}
\underset{q\rightarrow0}{\lim}q^{2}&\langle u_{i}(\bq)u_{j}(-\bq)\rangle=\label{eq:furpe}\\&=\underset{q\rightarrow0}{\lim}q^{2}\int Du\,u_{i}(\bq)u_{j}(-\bq)e^{-\int_{\bq'} f({\bf u}(\bq'))}\nonumber
\end{align}
Where $f({\bf u}(\bq))$ is the free energy density in Fourier space:
\begin{align}
   f(&{\bf u}(\bq))=\\&\frac{1}{2}\big{[}\mu(q^{2}\delta_{ij})u_{i}(\bq)u_{j}(-\bq)+(\lambda+\mu+\gamma)q_{i}q_{j}u_{i}(\bq)u_{j}(-\bq)\nonumber\\&-\gamma q_{x}q_{y}(u_{x}(\bq)u_{y}(-\bq)+u_{x}(-\bq)u_{y}(\bq))\big{]}\nonumber
\end{align}
Equation \eqref{eq:furpe} can be evaluated using 
\begin{align}
\langle u_{i}(q)u_{j}(-q)\rangle=\int Du\,u_{i}(q)u_{j}(-q)&e^{-\frac{1}{2}\int_{\bq'}[u^{\dagger}A u]}
\end{align}
where $A$ is a matrix of {\em renormalized} elastic constants,
\begin{align}
A 
=\left(\begin{array}{cc}
\mu q^{2}+(\lambda+\mu+\gamma)q_{x}^{2}&(\lambda+\mu)q_{x}q_{y}\\
(\lambda+\mu)q_{x}q_{y} & \mu q^{2}+(\lambda+\mu+\gamma)q_{y}^{2}
\end{array}\right)\nonumber  
\end{align}
and
\begin{align}
   u^\dagger=\begin{pmatrix}
u_x(-q)& u_y(-q) \
\end{pmatrix} ,\  u=\begin{pmatrix}u_x(q)\\
u_y(q)
\end{pmatrix}
\end{align}
So 
\begin{align}
\underset{q\rightarrow0}{\lim}q^{2}\langle u_{i}(q)u_{j}(-q)\rangle=\underset{q\rightarrow0}{\lim}q^{2}(A^{-1})_{ij} 
\end{align}
Then
\begin{small}
\begin{align}
G_{xx}(R)=&\\\ &\int_\bq\frac{\mu q^{2}+(\lambda+\mu+\gamma)q_{y}^{2}}{q^{4}\mu(2\mu+\lambda+\gamma)+2\gamma(\lambda+\mu+\gamma)q_{x}^{2}q_{y}^{2}}(e^{i{\bf q}\cdot {\bf R}}-1)\nonumber\\
G_{yy}(R)=&\nonumber\\ &\int_\bq\frac{\mu q^{2}+(\lambda+\mu+\gamma)q_{x}^{2}}{q^{4}\mu(2\mu+\lambda+\gamma)+2\gamma(\lambda+\mu+\gamma)q_{x}^{2}q_{y}^{2}}(e^{i{\bf q}\cdot {\bf R}}-1)\nonumber\\
G_{xy}(R)=&\nonumber\\ &\int_\bq\frac{-(\lambda+\mu)q_{x}q_{y}}{q^{4}\mu(2\mu+\lambda+\gamma)+2\gamma(\lambda+\mu+\gamma)q_{x}^{2}q_{y}^{2}}(e^{i{\bf q}\cdot {\bf R}}-1)\nonumber   
\end{align}
\end{small}
We are interested in the logarithmic dependence of these expressions, since this determines the power-law decay of the correlation function.  As discussed in Appendix ~\ref{sec:AppDerivationOfInteraction}, the coefficient of the logarithmic dependence is given by the angular average of the integrands.  For instance, the logarithmic dependence of $G_{xx}$ is obtained using,
\begin{align}
\int_0^{2\pi} \frac{d\theta}{2\pi} \frac{b+c\sin^2\theta}{1+d\cos^2\theta\sin^2\theta}=\frac{1}{2}\frac{b+c/2}{\sqrt{1+\frac{d}{4}}}\,.
\end{align}
$G_{yy}$ gives an identical contribution, whereas $G_{xy}$ averages to zero.  Plugging $G_{xx}(R)$ and $G_{yy}(R)$ into the correlation function we obtain
\begin{align}
C_{Q}(R)=e^{Q_{i}Q_{j}G_{ij}(R)}=\left(\frac{a}{R}\right)^{\eta_{Q}}  
\end{align}
With the exponent
\begin{align}
\eta_{Q}=\frac{\abs{Q}^{2}}{4\pi}\frac{(3\mu+\lambda+\gamma)}{\mu(2\mu+\lambda+\gamma)}\frac{1}{\sqrt{1+\frac{\gamma(\lambda+\mu+\gamma)}{2\mu(2\mu+\lambda+\gamma)}}}    
\end{align}
For the smallest reciprocal lattice vector $Q=\frac{2\pi}{a_0}$, we
get the decay exponent $\eta_0\equiv \eta_{\frac{2\pi}{a_0}}$, which equals
\begin{align}
\eta_{0}=\pi\frac{(3\mu+\lambda+\gamma)}{\mu(2\mu+\lambda+\gamma)}\frac{1}{\sqrt{1+\frac{\gamma(\lambda+\mu+\gamma)}{2\mu(2\mu+\lambda+\gamma)}}}\,.    
\end{align}
\subsubsection{Upper bound on the correlation length exponent}

We now give an upper bound on the correlation length exponent $\eta_0$.
We will denote the ratio $\frac{\gamma}{\mu}$ by $\phi\equiv\frac{\gamma}{\mu}$. From Eq.~\eqref{eq:mu} we have
\begin{align}
-2<\phi \label{eq:phi} 
\end{align}
We then express $K^{-1}$ and $\eta_{0}$ in terms of $\phi$, $\mu$ and $\sigma$:
\begin{align}
\eta_{0}=\pi\frac{1}{\mu}\left(\frac{3-\sigma+\phi}{2+\phi}\right)\frac{1}{\sqrt{1+\frac{\phi}{2}(1-\frac{1}{2+\phi}(1-\sigma))}}
\end{align}
\begin{align}
K^{-1}=\frac{1}{\mu(1+\sigma)(2+\phi)}\sqrt{1+\phi\frac{(1+\sigma)}{4}}
\end{align}

To get an upper bound on $\eta_0$ we use the fact that in the solid phase $K>16\pi$.  Therefore 
\begin{align}
\eta_0 < \frac{K\eta_{0}}{16\pi}\equiv \eta_m(\sigma,\phi)
\end{align}
where $\eta_m(\sigma,\phi)$ is given by
%
\begin{align}
 \eta_m(\sigma,\phi)\equiv\frac{(1+\sigma)(3-\sigma+\phi)}{16\sqrt{1+\frac{\phi}{2}\left(1-\frac{1-\sigma}{2+\phi}+\frac{1+\sigma}{2}-\frac{\phi(1-\sigma^2)}{4(2+\phi)}\right)+\frac{\phi^{2}}{2}\frac{1+\sigma}{4}}}
\end{align}
is the correlation length exponent at the melting temperature for a system with a given fixed value of $\sigma$ and $\phi$.

In the stable region, $-2<\phi$ and $-1<\sigma<1$, the function $\eta_m(\sigma,\phi)$
has a maximum of $1/4$, which is obtained when $\sigma=1$. Therefore,
\begin{align}
    \eta_0<\frac{1}{4}
\end{align}
is the bound for how quickly the positional correlations can decay in the solid phase.  We emphasize that this bound differs from that of an isotropic medium \cite{PhysRevB.19.2457},
\begin{align}
    \eta_0<\frac{1}{3} \,\,\,\, \, \, \mathrm{(isotropic)}
\end{align}
which is relevant for triangular lattice crystals.

\subsection{Orientational order in the tetratic phase}

In this section, we will compute the orientational decay exponent in the tetratic phase.  Our discussion is based on the analysis in \citep{PhysRevB.19.2457}.

\subsubsection{Residual orientational order}

In the tetratic phase, above the melting temperature $T_{m}$,
the effective Hamiltonian for the residual orientational order can
be written as \citep{PhysRevB.19.2457}
\begin{align}
H_{A}=\frac{1}{2}K_{A} \label{eq:bondangle}(T)\int\abs{\nabla\theta}^{2}d^{2}\br  
\end{align}
Where $K_{A}$ is the Frank constant and $\theta$ is the bond-angle field
angle, defined with the displacement field $\bf u$ by
\begin{align}
\theta=\frac{1}{2}\epsilon_{ij}\partial_{i}u_{j}
\end{align}
The Frank constant can be evaluated with:
\begin{align}
\frac{T}{K_{A}}=\underset{q\rightarrow0}{\lim}q^{2}\left\langle \widetilde\theta(\bq)\widetilde\theta(-\bq)\right\rangle     
\end{align}
Where $\widetilde\theta$ is the Fourier transform of $\theta$. In the background of a dislocation field in Fourier space \cite{PhysRevB.19.2457} \cite{cha95}:
\begin{align}
\theta_{s}(\bq)=-i\frac{\bq\cdot {\bf b }(\bq)}{q^{2}}    
\end{align}
The Frank constant is then:
\begin{align}
\frac{T}{K_{A}}=\underset{q\rightarrow0}{\lim}\frac{q_{i}q_{j}}{q^{2}}\left\langle b_{i}(q)b_{j}(-q)\right\rangle
\label{eq:FrankEvaluation}
\end{align}
Following Ref.~\onlinecite{PhysRevB.19.2457}, we compute $\left\langle b_{i}(\bq)b_{j}(-\bq)\right\rangle $ as a
response function with the dislocation Hamiltonian:
\begin{align}
\frac{H_{D}}{T}&=\frac{1}{2} \int_\bq\left(\frac{q^{2}\delta_{ij}-q_{i}q_{j}}{Yq^{4}+\alpha q_{x}^{2}q_{y}^{2}}+\frac{2E_{c}}{T}\delta_{ij}\right)b_{i}(\bq)b_{j}(-\bq) \\ &=\frac{1}{2}\int_\bq \left(P(\bq)\left(\delta_{ij}-\frac{q_{i}q_{j}}{q^2}\right)+\frac{2E_{c}}{T}\frac{q_{i}q_{j}}{q^2}\right)b_{i}(\bq)b_{j}(-\bq)\nonumber 
\end{align}
\noindent where the Burgers field is taken to be continuous.  Here, we defined 
\begin{align}
    P(\bq)=\frac{q^2}{Yq^{4}+\alpha q_{x}^{2}q_{y}^{2}}+\frac{2E_{c}}{T}
\end{align}
and $E_{c}$ is the core energy of a dislocation. Then,
\begin{align}
\left\langle b_{i}(q)b_{j}(-q)\right\rangle &=\int {\mathcal D}b\,b_{i}(q)b_{j}(-q)e^{-H_{D}/T}\\&=\frac{1}{P(\bq)}\left(\delta_{ij}-\frac{q_{i}q_{j}}{q^2}\right)+\frac{T}{2E_{c}}\frac{q_{i}q_{j}}{q^2}\nonumber
\end{align}
Substituting into Eq.~(\ref{eq:FrankEvaluation}), we obtain:
\begin{align}
\frac{T}{K_{A}}=\underset{q\rightarrow0}{\lim}\frac{q_{i}q_{j}}{q^{2}}\left\langle b_{i}(q)b_{j}(-q)\right\rangle=\frac{T}{2E_{c}}
\end{align}
Thus, the Frank constant is proportional to the core energy of a dislocation, just like in the triangular case.
\subsubsection{Critical exponent of the orientational order}
For a latice with $n$-fold rotational symmetry, 
the orientational correlation function is given by,
\begin{align}
    \left\langle e^{in(\theta(\textbf{R})-\theta(0)) }\right\rangle=e^{-\frac{n^2}{2}\left\langle{[\theta(\textbf{R})-\theta(0)]^2}\right\rangle}\sim R^{-\eta_n}
\end{align}
In the intermediate phase, this can be evaluated with the Hamiltonian ~\eqref{eq:bondangle}:
\begin{align}
    \left\langle{[\theta(\textbf{R})-\theta(0)]^2}\right\rangle=\frac{T}{\pi K_A}\ln\frac{R}{a}
\end{align}
Hence, the decay exponent for orientational correlations,
\begin{align}
    \eta_n=\frac{n^2T}{2\pi K_A}
\end{align}
is a function of the Frank constant.

At the transition from the intermediate phase to the liquid, $\eta_n$ obtains a universal value.  To see this, consider the effective Hamiltonian of disclinations in the intermediate phase,
\begin{align}
    H_{D}=-\frac{K_A}{4\pi}  \sum_{\br\ne \br^{\prime}} q(\br)q(\br^{\prime})\ln \abs{\frac{\br-\br^{\prime}}{a}}
\end{align}
where $q(\br)$ are charges given by the Frank angle of the disclinations,
\begin{align}
    q(\br)=\frac{2\pi}{n}s(\br).
\end{align}
Here we only consider fundamental disclinations, for which $s$ only takes the values $\pm 1$.
At the phase transition into the liquid, $T=T_i$, the Frank constant takes a universal value, $\frac{K_A(T_i)}{T_i}=\frac{2n^2}{\pi}$.  Hence, the decay exponent at the transition is 
\begin{align}
    \eta_n(T=T_i)=\frac{n^2}{2\pi}\frac{\pi}{2n^2}=\frac{1}{4}\,.
\end{align}
Note that although the disclinations of a square lattice have charge $\frac{\pi}{2}$ and not $\frac{\pi}{3}$ as in the triangular case, the critical exponent at melting is independent of $n$ and remains equal to $\frac{1}{4}$.

\section{Conclusions}
In this work, we extended the Kosterlitz-Thouless-Halperin-Nelson-Young (KTHNY) theory to the melting of square lattice solids, which are characterized by three independent elastic constants: the Young's modulus, the Poisson ratio, and an anisotropy parameter. By deriving the effective interactions between dislocations in this anisotropic medium, we showed that, despite some important differences from the triangular lattice case, the two-step melting scenario remains valid. 

We found that the transition from the solid phase proceeds through an intermediate tetratic phase, analogous to the hexatic phase in triangular lattices. The dislocation interactions are modified by anisotropy, leading to changes in both the universal quantities at the transition and in the decay of translational correlations. In particular, the bound on the translational correlation exponent $\eta_0$ is reduced from $\frac{1}{3}$ (in the isotropic case) to $\frac{1}{4}$ for the square lattice. Additionally, the Young's modulus by itself is no longer universal at the phase transition. Instead, we have a universal value for a certain combination of the Young's modulus and the anisotropy.

Finally, we demonstrated that orientational order persists in the tetratic phase with algebraically decaying correlations, characterized by a universal critical exponent $\eta_4=\frac{1}{4}$ at the tetratic-to-liquid transition, similar to the hexatic-to-liquid case.

It would be interesting to compare these predictions with experimental observations \cite{PhysRevE.76.040401,Walsh_2016,D4SM01377H} and numerical simulations \cite{article,Abutbul_2022,PhysRevResearch.7.L012034} of square lattice melting.  

Prior to concluding, we mention that the square lattice differs fundamentally from the triangular lattice in that it allows for two distinct types of disclinations \cite{PhysRevLett.111.025304}.  In particular, site-centered and plaquette-centered disclinations of the square lattice can be distinguished from each other by their topological properties.  This gives an entropic contribution per disclination, $\Delta S= T \ln 2$, which can be absorbed into the core energy.  Importantly, this distinction does not affect the long-range interaction between disclinations, and therefore we do not expect it to play a role in the melting transitions.

\acknowledgments

We are grateful to Leo Radzihovsky, Ari Turner, and Zack Weinstein for useful discussions.  DP acknowledges funding from the Israel Science Foundation (ISF) under grant no.~2005/23.

\appendix

\section{Derivation of the interaction between dislocations}
\label{sec:AppDerivationOfInteraction}
Here we develop the interaction between dislocations. We start with a pair of dislocations with opposite Burgers vectors, ${\bf b}_1=\hat{x}$ and ${\bf b}_2=-\hat{x}$, separated by a displacement ${\bf R}$.  The interaction energy is then given by Eq.~\ref{eq:twodisloc}:
\begin{align}
 E&=\frac{Y a_0^2}{2}\int_\bq\frac{q^{2}_y\,\left(2-2e^{i{\bf R}\cdot \bq}\right)}{q^{4}+\zeta q_{x}^{2}q_{y}^{2}}  \label{eq:twodislocap}\\&=\frac{Y a_0^2}{2}\int_\bq\frac{g(\theta)}{q^2}\left(2-2e^{i\cos(\theta-\phi) Rq }\right)\nonumber
\end{align}
After the change of variables $\theta\rightarrow\theta+\phi$ and using \begin{align} 
g(\theta)=\sum_{n=0}{h_{2n} \cos(2n\theta)}\, ,
\end{align} ~\eqref{eq:twodislocap} becomes:
\begin{align}
 & E=Y a_0^2\int_\bq\frac{g(\theta+\phi)}{q^2}\left(1-e^{i\cos\theta Rq }\right)\\
 &=\frac{Y a_0^2}{2\pi}\int_{\frac{R}{L}}^{\frac{R}{a}}\frac{du}{u}\int_0^{2\pi}\frac{d\theta}{2\pi}\sum_{n=0}^\infty{h_{2n} \cos(2n(\theta+\phi))}\left(1-e^{i\cos\theta u}\right)\nonumber
\end{align}
where $L$ is the size of the system, $a$ is the dislocation core radius and we used the variable change $u=Rq$. We next use
\begin{align}
   \cos(2n(\theta+\phi))=\cos(2n\theta) \cos(2n\phi)-\sin(2n\theta)\sin(2n\phi)\nonumber
\end{align}
and the fact that $\sin(2n\theta)$ is antisymetric in $\theta$, so it falls off in the angular integration. The $n$-th term in the sum will become the $2n$-th Bessel function $J_{2n}(u)$ with the definition
\begin{align}
    \frac{1}{2\pi}\int_0^{2\pi}d\theta\cos(2n\theta)e^{iu\cos\theta}=(-1)^nJ_{2n}(u)
\end{align}
In addition, only the $n=0$ term survives the angular integration over $1$. We get:
\begin{align}   
E&=\frac{Y a_0^2}{2\pi}\int_{\frac{R}{L}}^{\frac{R}{a}}\frac{du}{u}\left(h_0-\sum_{n=0}^\infty(-1)^nh_{2n}\cos(2n\phi)J_{2n}(u)\right)\nonumber\\
&=\frac{Y a_0^2}{2\pi}\left(\int_{\frac{R}{L}}^{\frac{R}{a}}\frac{du}{u}h_{0}(1-J_{0}(u))\right.\\&\qquad \qquad\qquad \left.-\sum_{n=1}^\infty(-1)^nh_{2n}\cos(2n\phi)\int_{0}^{\infty}du\frac{J_{2n}(u)}{u}\right)\, \nonumber 
\end{align}
The first part is independent of the relative orientation between the dislocations and it gives a logarithmic dependence in $R$
\begin{align}
    \frac{Y a_0^2}{2\pi}h_0&\int_{\frac{R}{L}}^{\frac{R}{a}}du\frac{1-J_0(u)}{u}=\frac{Y a_0^2}{2\pi}h_0\left(\ln\frac{L}{a}+\ln\frac{R}{L}\right)\nonumber\\&=\frac{Y a_0^2}{4\pi}\frac{1}{\sqrt{1+\frac{\zeta}{4}}}\ln\frac{R}{a}\,.
\end{align}
Thus, the logarithmic dependence in the interaction comes from $h_0$, i.e. from the angular average of $g(\theta)$.  The second term, with $n\ge 1$, yields converging integrals in both the $u\to 0$ and $u\to\infty$ limits and is therefore independent of $R$ provided that $R/a\gg1$ and $R/L\ll1$.  This term contains the angular dependence as a sum of even harmonics of $\phi$. By using
\begin{align}
\int_{0}^{\infty}du\frac{J_{2n}(u)}{u}=\frac{1}{2n}   
\end{align}
the overall angular interaction is:
\begin{align}
\sum_{n=1}(-1)&^nh_{2n}\cos(2n\phi)\int_{0}^{\infty}du\frac{J_{2n}(u)}{u}\\&=\sum_{n=1}(-1)^n\frac{h_{2n}}{2n}\cos(2n\phi)\nonumber
\end{align}

We now turn to the interaction between dislocations with perpendicular Burgers vectors, ${\bf b}_1=\hat{x}$ and ${\bf b}_2=\hat{y}$.  Returning to Eq. \ref{eq:dislocE} and plugging the perpendicular dislocations we get
\begin{align}
E&=\frac{Y a_0^2}{2}\int\frac{d^{2}{\bf q}}{(2\pi)^{2}}\frac{q^{2}-2q_xq_ye^{i{\bf R}\cdot \bq}}{q^{4}+\zeta q_{x}^{2}q_{y}^{2}}\ . \label{eq:twoperpdisloc}
\end{align}
The first part produces a logarithmically-divergent self-interaction.  However, we assume that these dislocations are part of a neutral ensemble.  Then, this term combines with the interaction with opposite Burgers vector dislocations to yield a finite energy, which is the pairwise interaction between opposite charges as computed in Eq.~(\ref{eq:twodislocap}).  
Hence, only the second term must be considered.  It can be written as
\begin{align}
\widetilde{W}(\phi)=-Y a_0^2\int\frac{d^{2}{\bf q}}{(2\pi)^{2}}\frac{\widetilde{g}(\theta)}{q^2}e^{i\cos(\theta-\phi) Rq}
\end{align}
with 
\begin{equation}
    \widetilde{g}(\theta)=\frac{\sin\theta\cos\theta}{1+\zeta \cos^2\theta\sin^2\theta}=\frac{\frac{1}{2}\sin(2\theta)}{1+\zeta \cos^2\theta\sin^2\theta}\,.
\end{equation}
Now $\widetilde{g}(\theta)$ is odd and a periodic function of $\theta$ with period $\pi$. It can be decomposed as:
\begin{align} 
\tilde{g}(\theta)=\sum_{n=1}{\tilde{h}_{2n} \sin(2n\theta)} \label{eq:harmonic3}
\end{align}
with
\begin{align}
  \widetilde{h}_{2n}=\int_0^{2\pi}\frac{d\theta}{2\pi}2\sin(2n\theta) \widetilde{g}(\theta) \ .  
\end{align}
We note that $\tilde{g}(\theta)$ is even about $\theta=\pi/2$, which implies that $\tilde{h}_{2n}$ vanishes for even values of $n$. Following the same procedure as above, we obtain 
\begin{align}
   \widetilde{W}(\phi)=Y a_0^2\sum_{n=1}\frac{\widetilde{h}_{2n}}{4\pi n}\sin(2n\phi)
\end{align}
which is the interaction energy between perpendicular dislocations.  In this case, there is no dependence on the distance between the dislocations, and the interaction energy is purely angular.

\section{Derivation of the inverse elastic tensor}
\label{app:DerivationInverseTensor}

Here we find the inverse of the elastic tensor. By its definition, the tensor $C_{ij,kl}$ is symmetric with respect to the exchange of the indices $i\leftrightarrow j$, of $k \leftrightarrow l$, and also of the index pairs $(ij)\leftrightarrow(kl)$. We wish to find the inverse elastic tensor, with the definition 
\begin{align}
C_{ij,kl}C_{kl,mn}^{-1}=\frac{\delta_{im}\delta_{jn}+\delta_{in}\delta_{jm}}{2}    
\end{align}
We proceed by defining the tensors
\begin{align}
A_{ij,kl}&\equiv\delta_{ik}\delta_{jl}+\delta_{il}\delta_{jk}\nonumber\\
B_{ij,kl}&\equiv\delta_{ij}\delta_{kl}\\
D_{ij,kl}&\equiv\delta_{i1}\delta_{j1}\delta_{k1}\delta_{l1}+\delta_{i2}\delta_{j2}\delta_{k2}\delta_{l2}\nonumber    
\end{align}
These tensors satisfy the algebra
\begin{align}
\hat{A}\hat{A}=2 \hat{A}, \quad \hat{A}\hat{B}=2 \hat{B},\quad \hat{A}\hat{D}=2 \hat{D}\label{eq:alg}\\ 
\hat{B}\hat{B}=2 \hat{B}, \quad \hat{B}\hat{D}=2 \hat{B},\quad \hat{D}\hat{D}= \hat{D}\nonumber   
\end{align}
Where $\hat{A}\hat{A}=2\hat{A}$ is a shorthand notation for $A_{ij,kl}A_{kl,mn}=2A_{ij,mn}$.  To find $\hat{C}^{-1}$, we expand $\hat{C}$ and $\hat{C}^{-1}$ as linear combinations of these tensors:
\begin{align}
\hat{C}=\mu\hat{A}+\lambda\hat{B}+\gamma\hat{D}\\
\hat{C}^{-1}=a\hat{A}+b\hat{B}+c\hat{D}\nonumber
\end{align}
Then, the coefficients $a$, $b$, and $c$ are found by requiring
\begin{align}
\left(\mu\hat{A}+\lambda\hat{B}+\gamma\hat{D}\right)\left(a\hat{A}+b\hat{B}+c\hat{D}\right)=\frac{\hat{A}}{2}\,.  
\end{align}
Using the algebra in Eq.~\eqref{eq:alg} we obtain the inverse elastic tensor:
\begin{align}
C_{ij,kl}^{-1} & =\frac{1}{4\mu}(\delta_{ik}\delta_{jl}+\delta_{il}\delta_{jk})-\frac{\lambda}{(2\mu+\gamma)(2\mu+2\lambda+\gamma)}\delta_{ij}\delta_{kl}\nonumber\\
 & -\frac{\gamma}{2\mu(2\mu+\gamma)}(\delta_{i1}\delta_{j1}\delta_{k1}\delta_{l1}+\delta_{i2}\delta_{j2}\delta_{k2}\delta_{l2})
\end{align}

\medskip

\end{document}